\documentclass[gsifonts,twoside,12pt]{gsipaper}
\usepackage{a4wide,gsiindex,helvet}
\usepackage[english]{babel}
\usepackage{amsfonts,amssymb,verbatim,graphicx,float}
\usepackage{epsfig}

\usepackage[utf8]{inputenc}
\usepackage{pstricks}
\usepackage{pst-plot}

\newcommand{\eqnref}[1]{Eq.~(\ref{#1})}

\newcommand{\figref}[1]{Fig.~\ref{#1}}

\newcommand{\tabref}[1]{Table~\ref{#1}}

\def\slashchar#1{\setbox0=\hbox{$#1$}
   \dimen0=\wd0 \setbox1=\hbox{/} \dimen1=\wd1
   \ifdim\dimen0>\dimen1 \rlap{\hbox to \dimen0{\hfil/\hfil}} #1
   \else  \rlap{\hbox to \dimen1{\hfil$#1$\hfil}} / \fi}

\begin{document}

\title{Jet-Evolution in the Quark-Gluon Plasma from RHIC to the LHC
}

\date{\today}

\abstract{The observed suppression of high-$p_\perp$ hadrons allows different explanations. We discuss two possible scenarios: In scenario 1, parton energy loss from scattering in the hot medium is complemented by final state interactions in the resonance matter. Scenario 2 has an enhanced transport parameter $\hat q$ which is fitted to RHIC data. For LHC, the two scenarios lead to very different predictions for the nuclear modification factor of hadrons. In addition, jet reconstruction allows more specific tests of the mechanisms responsible for jet quenching. We calculate the distribution of partons inside a jet and find different results for the two scenarios.  }

\author{S.~Domdey$^a$, B.Z.~Kopeliovich$^{a,b}$, H.J.~Pirner$^{a}$}

\address{
a) Institute for Theoretical Physics, 
University of Heidelberg,
Germany\\
b) Departamento de F\'{\i}sica,
Universidad T\'ecnica Federico Santa Mar\'{\i}a;\\
Instituto de Estudios Avanzados en Ciencias e Ingenier\'{\i}a; \\
Centro Cient\'ifico-Tecnol\'ogico de Valpara\'iso,
Casilla 110-V, Valpara\'iso, Chile
}

\maketitle

%
%

\section{Introduction}
\label{sec:intro}
Investigations of the hot medium produced in heavy-ion collisions largely suffer from both its short life time and the small production rate of hard probes. Proceeding from experiments at RHIC to LHC, these features will be significantly improved. In particular, the increase in energy at LHC will lead to a higher temperature compared to RHIC which corresponds to a longer life time. Secondly, full jet reconstruction provides more differential information on the mechanisms of final state interactions compared to spectra of leading hadrons.

The large suppression of leading hadrons as observed by PHENIX and STAR coll. \cite{Adler:2003qi} (see also \cite{Arsene:2004fa}) has triggered large interest in final state interactions in the hot medium produced in heavy-ion collisions. Theoretical models have focussed mostly on the energy loss of fast partons during their propagation through the dense medium. Many discussions of the range of mechanisms at work at RHIC and possibly at LHC can be found in the literature (see for example \cite{Majumder:2010qh}).



In the picture of partonic energy loss, hadronization takes place outside of the hot zone and is not affected by the medium.
Special features may develop due to the nonabelian nature of QCD. 
Coherent radiative energy loss may be important due to the radiated gluon interacting with the medium.
This explanation corresponds to a less pronounced suppression of heavy charmed quark/hadrons in contrast to experimental results \cite{Adare:2006nq}. 

Gluon radiation undoubtedly plays an important role for parton evolution at RHIC and the LHC. Typical virtualities of
partons are around $Q=20 $ GeV  and $Q=100 $ GeV, respectively. The resulting 
parton showers extend over several fm  and evolve in the medium. Therefore,
a separation of this shower from the propagation of partons in the medium seems to
be unrealistic. From our perspective, however,
there are no strong indications for coherence yet, because of the finite size of the medium which blurs the characteristic
dependence of the energy loss on the square of the path length \cite{Baier:1996kr,wang,gyulassy,wiedemann,Zakharov:1998sv}.  

In our view, it is
pragmatic to explore parton evolution interleaving 
radiation and interaction with the medium incoherently.
Once the parton dynamics has been settled, measurements of the high-$p_\perp$
hadrons and/or jets can be used as a tool to investigate the properties of
the quark gluon plasma. The 
further hypothesis of a strongly interacting quark gluon plasma (sQGP) \cite{Shuryak:2004cy}
can be tested. This idea underlies models with large energy loss parameters $\hat q$, but has not  received so much backing from lattice simulations where
the quark gluon coupling constant was monitored in the region
$0.5<\alpha_s<1$ at most \cite{Petreczky:2007bn}.

A typical shower at RHIC evolves from a high virtuality around $Q=20 $ GeV to a 
low perturbative scale $Q_0=1.5$ GeV and lasts about 
$\tau_{\rm evo}=2 $ fm. In comparison, for longitudinal expansion with an initial temperature of $T_0=300$ MeV at $\tau_0=
0.5$ fm, one may estimate a plasma lifetime of $\tau_c= 3.3
$ fm [For details on these estimates, see \eqnref{tauc} and \eqnref{tauevo}.]. From these  two time estimates, we find 
\begin{equation}
\tau_{\rm evo}\approx \tau_c.
\end{equation}
The nonperturbative part of hadronization involves the decay of the prehadrons or
resonances at the preconfinement scale $Q_0$ where e.g. the vector meson resonances decay into $4-5$ pions.
Therefore, it is very probable at RHIC that at the end of the evolution resonances interact with the remainder of the quark-gluon plasma, i.e. with hadronic resonance matter \cite{Karsch:2003vd}.
For $T\leq T_c$, the equation of state of hadronic matter can well be described by a hadron resonance gas with an initial density $n_{\rm res}\approx T_c^3$ \cite{Gerber:1988tt}. For our purposes, we neglect the slow cool-down of resonance matter below $T_c$. Such an interaction 
may be described by a hadronic theory with cross
sections slightly larger than hadronic cross sections in
vacuum. In general, hadronic resonances have radii which are larger than hadronic ground states. Therefore, one expects that $q\bar q$-resonances have larger cross section than the $q\bar q$-ground states. Because of these large cross sections, newly formed prehadrons are absorbed. We think that final state effects of
these resonances with the hot resonance gas play a decisive role in the
observed suppression of hadrons in RHIC experiments.

We therefore would like to advocate a comparison of
two scenarios. 
\begin{itemize}
\item Scenario 1 uses an estimate of $\hat q = 0.5$ GeV$^2$/fm for the jet transport parameter at RHIC.  This scenario underestimates the suppression when only parton energy loss is taken into account.
Additionally, we consider the absorption of the prehadrons in the resonance gas, leading to a similar suppression as observed at RHIC.
\item Scenario 2 considers large parton energy loss solely and tunes up this value to $\hat q=4$ GeV$^2$/fm in order to accomodate RHIC data. 
\end{itemize} 
Phenomenologically, we propose to test both scenarios on LHC data. For this purpose, we extrapolate both scenarios  to Pb+Pb collisions at LHC with $\sqrt{s}=5.5$ TeV, a virtuality of $Q=100 $ GeV and an initial temperature of $T_0=500$ MeV.
In the LHC case, the average time for the evolution from the initial high virtuality to a hadronic scale $Q_0$ is much longer than the plasma life time and the size of the hot interaction zone, 
\begin{equation}
\tau_{\rm evo}>\tau_c.
\end{equation}
Therefore we expect that resonances do not play an important role there. 

Another important feature of LHC experiments is the possibility of full jet reconstruction up to very small momenta. Therefore we will also make predictions
for jet compositions down to very small
momentum fractions which govern the jet multiplicity. In both scenarios,  the
respective temperatures and virtualities control  the parametric 
dependence of the input data to the modified DGLAP equations we use. The main uncertainty concerns the question whether final state absorption is important. Should we describe RHIC physics in the pure energy loss scenario 2 or in the mixed 
scenario 1, where nonperturbative features of resonance interactions occur together with perturbative parton interactions during hadronization? We claim that LHC will tell which scenario is preferred with its 
forthcoming experimental results at the end of this year.


\section{Formalism and Modified Evolution Equations}
Hard processes in a heavy-ion collision lead to the production of
partons with high virtuality and high energy.  These partons will successively radiate gluons
 to reduce their virtuality and become on mass-shell. This
leads to a parton shower and scaling violations in the vacuum jet
fragmentation functions $D_i^v(x,Q^2)$ described by the DGLAP equations \cite{DGLAP}
\begin{equation}
\frac{\partial}{\partial\ln
Q^2}D_i^v(x,Q^2)=\frac{\alpha_s(Q^2)}{2\pi} \int_x^1 \frac{\mbox{d}
z}{z} P_{ji}(z) D_j^v\left(\frac{x}{z},Q^2 \right).
\end{equation}
In Ref.~\cite{Domdey:2008gp}, some of the authors have developed a model for the parton cascade that includes
scattering off the partons in the QGP. In this model, we modified the DGLAP equations
by an additional scattering term.

In this section, we repeat the formalism for our further considerations.
We consider both quarks and gluons as hadronizing partons. Indices on the fragmentation functions and
the splitting functions indicate the respective parton species.

The formation of a parton shower does not happen instantaneously, but
requires a certain time. The relevant time is
the time a virtual state needs to evolve in virtuality from $Q^2$ to
$Q^2+\mbox{d} Q^2$,
\begin{equation}\label{dtau}
\mbox{d}\tau= \frac{E}{Q^2}\frac{\mbox{d}Q^2}{Q^2},
\end{equation}
where $E$ is the energy of the parton and $Q$ its virtual
mass.
The time a parton needs to reduce its virtuality from the starting
scale $Q_{ i} \simeq E$ to the hadronization
scale $Q_0 \simeq 1$ GeV can then be estimated as $\tau_{\rm evo} = \int
\mbox{d}\tau \approx \frac{E}{Q_0^2}-\frac{1}{E}$. Thus, even though a
high energy parton with large virtuality will reduce its virtuality
rapidly, the overall lifetime is considerable and of the order of
several fermi. 

Furthermore, the plasma life time $\tau_{c}\simeq\tau_0(T_0/T_c)^3$ (corresponding to the cool-down from the
initial temperature to the critical temperature) may be long and of similar magnitude. Consequently, the parton shower overlaps in
time with the plasma phase. Therefore, splitting and scattering
processes have to be treated in a common framework.
During the time a fast parton reduces its virtuality by building up the parton
shower, it experiences scatterings with gluons of thermal mass
$m_s$ in the plasma. Therefore, we include a scattering term $S(x,Q^2)$ into the DGLAP evolution for the fragmentation function $D^m(x,Q^2)$  in the medium.
The resulting evolution equation combines radiation and
scattering:
\begin{equation}\label{mod-ev}
\frac{ \partial\, D^m(x,Q^2)}{\partial\, \ln{Q^2}}
=\frac{\alpha_s(Q^2)}{2 \pi} \int_{x}^1 \frac{\mbox{d}z}{z} P(z)
D^m\left(\frac{x}{z},Q^2\right)+ S(x,Q^2).
\end{equation}

For the construction of the scattering term we
estimate the relative importance of scatterings with the help of the
scattering mean free path $\lambda=(n \sigma)^{-1}$ by
\begin{equation}\label{scatt-prob}
\frac{\mbox{d}\tau}{\mbox{d}\lambda}=\frac{E_{\rm in}}
{Q^2} n \frac{\mbox{d}\sigma}{\mbox{d}  q_\perp^2} \mbox{d}  q_\perp^2 \frac{\mbox{d}Q^2}{Q^2}.
\end{equation}
Two contributions to scattering have to be
taken into account: A gain term for scattering from a higher
energy fraction to the given energy fraction $x$ and a loss term for
scattering from $x$ to lower energy fractions.  Consequently, the
energy $E_{\rm in}$ of the incoming parton (before the scattering
takes place) is different in the gain and the loss term (c.f. \eqnref{Sfirst}).

Scattering makes the fast parton lose energy, which is absorbed as
recoil energy $q_\perp^2/(2m_{s})$ by the plasma parton.
In our treatment, we allow only soft scattering in the $t$-channel, i.e. small relative
transverse momentum $q_\perp^2 \sim m_D^2$, such that the change of longitudinal momentum fraction $\Delta x= y-x =
q_\perp^2/(2 m_{ s} E)$ is small. The Debye mass $m_D$ of the gluon and the thermal mass $m_s$ are related as $2m_s^2=m_D^2$ \cite{kapusta} and $m_D\approx 3T$ \cite{Maezawa:2009nd}. In contrast to splitting
processes, which lead to a multiplicative change of the energy
fraction of the parton, we model soft scattering processes by shifts
to smaller energy fractions.

Scattering is included in the evolution equation in a similar way as
radiation. The ``scattering probability'' \eqnref{scatt-prob} is
folded with the fragmentation functions and gain and loss term are
subtracted from each other. The additional scattering term in the evolution equation has the following form
\begin{eqnarray}\label{Sfirst}
 S(x, Q^2)&= &\frac{ E}{Q^2} \bar n \int _x^1 \mbox{d}w \int \mbox{d}  q_\perp^2
  \frac{\mbox{d}\bar \sigma}{\mbox{d} q_\perp^2} \left( w D^m(w,Q^2)-x D^m(x,Q^2)\right)\nonumber\\
&& \hspace{2cm}\times \delta\left(w- x-\frac{ q_\perp^2}{2 m_s E}\right).
\end{eqnarray}
We then expand in powers of $ q_\perp^2/(2m_s E)$
and drop terms of second order and higher:
\begin{eqnarray}\label{Sterm}
S(x, Q^2)&=&\hspace{-0.2cm} \frac{ E}{Q^2} \bar n \int \mbox{d} q_\perp^2
  \frac{\mbox{d}\bar \sigma}{\mbox{d} q_\perp^2}
  \left[\left(x+\frac{q_\perp^2}{2m_s E}\right) D\left(x+\frac{
  q_\perp^2}{2m_s E},Q^2\right)-x D(x,Q^2)\right]\nonumber\\ &\simeq
  &\frac{ \bar n}{2 m_s Q^2} \int \mbox{d} q_\perp^2
  \frac{\mbox{d}\bar \sigma}{\mbox{d} q_\perp^2} q_\perp^2 \left(D(x,Q^2)+x
  \frac{\partial D}{\partial x}(x,Q^2)\right)\nonumber\\ &\simeq
  &\frac{ \bar n\bar \sigma \langle q_\perp^2 \rangle}{2 m_s Q^2}
  \left(D(x,Q^2)+x \frac{\partial D}{\partial x}(x,Q^2)\right)
  \hspace{-1cm}.
\end{eqnarray}
Such an expansion is meaningful for small momentum transfer, $
q_\perp^2/(2m_s E)\ll 1$ and $x+ q_\perp^2/(2m_s E) < 1$ as we expect them 
for small momentum transfers
$ q_\perp^2 \sim m_D^2$ and large jet energies.  Remarkably, the
scattering term does not depend on the jet energy $E$ in this
approximation.

In the expression above, $\bar n \, \frac{\mbox{d}\bar \sigma}{\mbox{d} q_\perp^2}$ denotes the weighted sum of possible scattering channels. Depending on the incident parton, these can be
 quark-quark, quark-gluon and gluon-gluon scatterings:
\begin{eqnarray}
\bar n \frac{\mbox{d} \bar \sigma}{\mbox{d} q_\perp^2}= \frac{16}{\pi^2}\zeta(3)T^3 \frac{2\pi\alpha_s(Q^2)^2}{ (q_\perp^2 +m_D^2)^2}\times 
\left\{
\begin{array}{cc}
(1+\frac78)=\frac{15}{8}& \mbox{incident quark}\\
(\frac94+\frac{63}{32})=\frac{135}{32}& \mbox{incident gluon}\\
\end{array} 
\right. .
\end{eqnarray} 
Interestingly, the quantity appearing in the numerator in \eqnref{Sterm} is the jet
transport parameter \cite{Baier:1996kr}
\begin{equation}
\hat q \simeq \bar n \bar \sigma \langle q_\perp^2 \rangle,
\end{equation}
which describes the mean acquired transverse momentum per unit
length (For a recent discussion of $\hat q$ see \cite{Bass:2008rv}). 

For an estimate of the numerical value of $\hat q$ associated with this expression, we use  
\begin{equation}\label{qhatval}
 \langle q_\perp^2 \rangle \simeq m_D^2,\qquad \hat q\simeq \frac{16}{\pi^2}\zeta(3)T^3 \times 2\pi\alpha_s^2(Q_0^2) \times
\left\{
\begin{array}{cc}
(1+\frac78)=\frac{15}{8}& \mbox{incident quark}\\
(\frac94+\frac{63}{32})=\frac{135}{32}& \mbox{incident gluon}\\
\end{array} 
\right. .
\end{equation}
The scattering term is most relevant at small virtualities $Q\simeq
Q_0$ and consequently we use  $\alpha_s(Q_0)$ to
arrive at an upper boundary for $\hat q$. Explicitly, \eqnref{qhatval} gives $\hat q\simeq 0.5$ GeV$^2$/fm for a temperature $T=0.3$
GeV for RHIC and $\hat q=5.2$ GeV$^2$/fm for $T=0.5$ GeV
corresponding to LHC. To fit the experimental data with parton energy
loss exclusively, we introduce an enhancement factor $K$ in the scattering term.
Then the scattering term reads
\begin{equation}\label{Sf}
S(x,Q^2)= K\frac{ \bar n \bar \sigma \langle q_\perp^2 \rangle}{2 m_s Q^2}
\left(D^m(x,Q^2)+x \frac{\partial D^m}{\partial x}(x,Q^2)\right).
\end{equation}
The value of $\hat q=0.50$ GeV$^2$/fm ($K=1$) has to be viewed in the context
of scenario 1 discussed in the introduction which combines parton energy loss with
resonance absorption in the fading plasma. Pure parton energy loss in scenario 1 gives a too weak
suppression factor to explain experimental data from central Au+Au-collisions at RHIC.

In scenario 2 without final state absorption, we need roughly $K=8$ or $\hat q\simeq 4.0$ GeV$^2$/fm to reproduce the observed nuclear modification factor. If parton energy loss dynamics is the same at LHC, this estimate would lead to a value as large as $\hat q=41.9$
GeV$^2$/fm (for $K=8$). This increase orginates equally from the
higher temperature $T=0.5$ GeV and the larger scattering cross section of gluons. The values for $\hat q$ for both scenarios
are summarized in \tabref{hatq}.

In our calculation above, we have introduced $K$ as an artifical enhancement parameter. The physical motivation for this soft enhancement factor may come from our estimate of the scattering cross section. The perturbative running coupling may  underestimate the soft scattering cross section of the parton in the plasma \cite{Braun:2006vd} if one assumes that the temperature defines the relevant scale. More precisely, evaluating the running coupling at $\pi T$ instead of $Q_0$ may lead to an enhancement factor of $K\leq 3$ for RHIC temperature. A factor of $K=8$ as in scenario 2 cannot be explained. 
In contrast, for the higher temperature at LHC the scales $\pi T$ and $Q_0$ are approximately equal which makes an enhancement factor $K$ unlikely.

\renewcommand\arraystretch{1.3}
\begin{table}
\begin{center}
\begin{tabular}{|c||c|c|}\hline
 $\hat q$ [GeV$^2$/fm] & $T=0.3$ GeV & $T=0.5$ GeV \\ \hline\hline
Scenario 1 (parton energy loss + absorption), $K=1$ & 0.5  & 5.2  \\
Scenario 2 (large parton energy loss), $K=8$ & 4.0 & 41.9\\ \hline
\end{tabular}
\end{center}
\caption{\label{hatq} Relevant values of the jet transport parameter $\hat q$ in the two scenarios for $T=0.3$ GeV (incident quark) and $T=0.5$ GeV (incident gluon). The effective $K$-factor allows possible enhancements of the energy loss parameter $\hat q$.}
\end{table}

Up to now, the theoretical formalism has focussed on processes at the parton level. In order to compare with observable hadrons, one has to
account for hadronization.

\section{Modified Fragmentation Functions $D_u^\pi,D_g^\pi$ in the Quark-Gluon Plasma}
\label{mod-ff}
For a solution of the modified DGLAP equation (\ref{mod-ev}), we evolve a given fragmentation function $D(x,Q_0^2)$
 to the ultraviolet scale $Q_{\rm max}$, i.e. we numerically follow a
way opposite to the shower evolution itself. Furthermore, we assume that the fragmentation function at
the infrared cut-off scale $Q_0$ is unchanged and use  standard vacuum fragmentation functions as
initial condition. We use the AKK parametrization \cite{Albino:2005me} for the fragmentation of
$u$-quarks and gluons into pions
\begin{equation}\label{AKK}
D_{ u}^{v,\pi}(x,Q_0^2)=0.447\,x^{-1.58}\,(1-x)^{1.01},\qquad D_{ g}^{v,\pi}(x,Q_0^2)=429\,x^{2.00}\,(1-x)^{5.82}
\end{equation}
at the scale $Q_0^2 = 2 \mbox{ GeV}^2$ in the evolution equations. The
numerical solution of the modified DGLAP evolution equation
(\ref{mod-ev}) (with and without scattering term) is calculated
with the Runge-Kutta method of 4th order.

\begin{figure}[!t]
\centering
\includegraphics[width=0.8\textwidth]{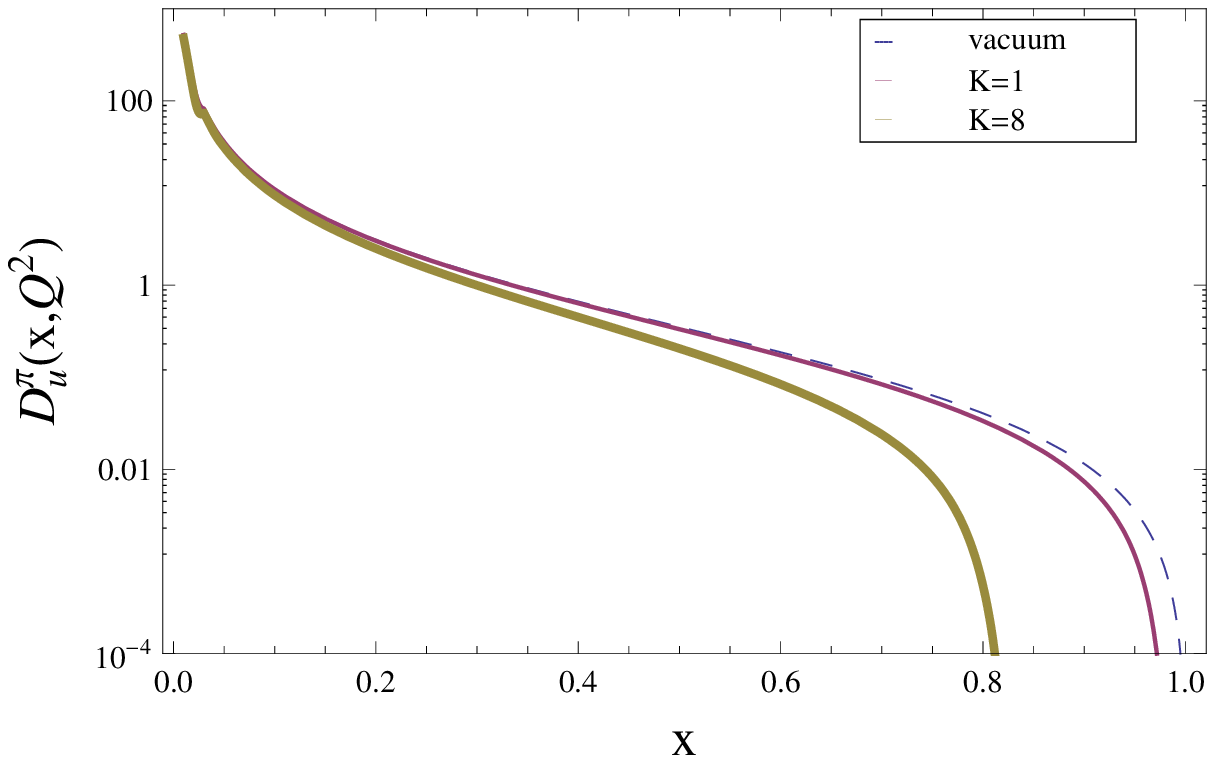}
\caption{\label{ff-rhic} Fragmentation functions $D_{u}^\pi (x,Q^2)$
  in medium and vacuum as a function of energy fraction $x$. The plot
  compares the results for vacuum evolution (  dashed curve) to
  medium-modified evolution with $K=1$ (full drawn thin curve) and $K=8$
  (full drawn thick curve). For the calculation, we have used $Q_{\rm max}=20$ GeV
  and $T=0.3$ GeV in \eqnref{mod-ev}.  }
\end{figure}

\begin{figure}
\centering
\includegraphics[width=0.8\textwidth]{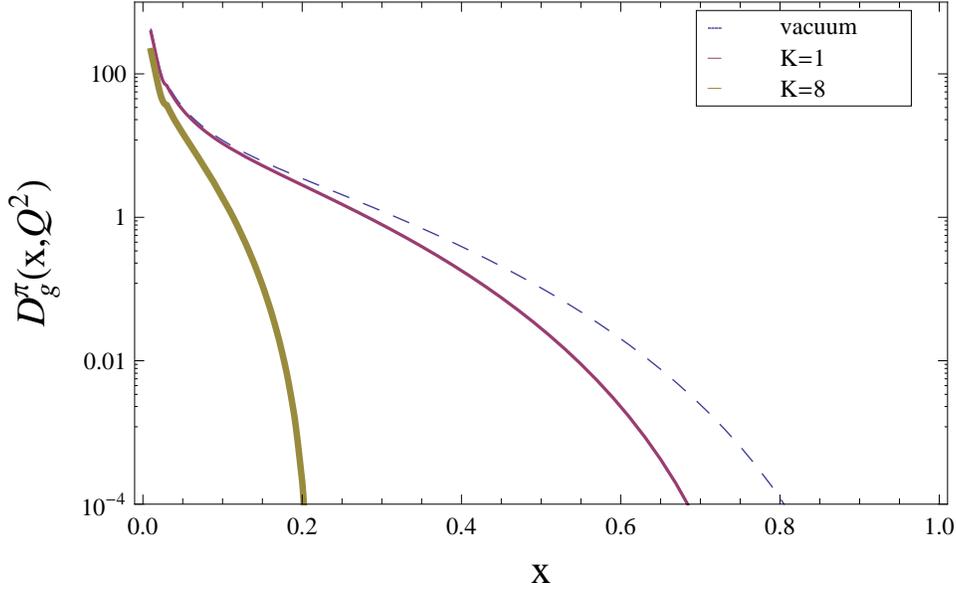}
\caption{\label{ff-lhc} Fragmentation functions $D_{g}^\pi (x,Q^2)$ in
  medium and vacuum as a function of energy fraction $x$. The plot
  compares the results for vacuum evolution (  dashed curve) to
  medium-modified evolution with $K=1$ (full drawn thin curve) and $K=8$
  (full drawn thick curve). For the calculation, we have used $Q_{\rm max}=100$ GeV
  and $T=0.5$ GeV in \eqnref{mod-ev}.  }
\end{figure}


At RHIC the fragmentation of quarks is the main source for pion production.  To investigate the modification of the $u$-quark to pion fragmentation
function in the medium at RHIC, we have solved \eqnref{mod-ev} for $Q_{\rm max}=20$ GeV
and $T=0.3$ GeV. Two different values for the enhancement parameter $K$ are used: $K=1$ and $K=8$. 

The results for the gluon to pion fragmentation function are shown in \figref{ff-rhic}.  The overall behaviour of the medium-modified fragmentation function with $Q^2$ is similar to the vacuum results.  The scattering term acts in a similar manner as the
splitting term and ``transports'' partons from large to small energy fractions.  The suppression of leading particles is strongest at large $x$.  Following naive expectations, the suppression of the medium fragmentation function becomes stronger for larger $K$.

For LHC, gluon fragmentation is more important than quark fragmentation. The resulting gluon-to-pion fragmentation functions for $Q_{\rm max}=100$ GeV and $T=0.5$ GeV are shown in \figref{ff-lhc} for both the vacuum case and for the case of a plasma with $K=1$ and $K=8$.  Note that the higher temperature at LHC and the larger color charge of the gluon lead to much larger values of
$\hat q$ (see \tabref{hatq}) although the microscopic dynamics
(i.e. energy loss of a fast parton) is unchanged.
The scattering term drastically reduces the number of partons with large
$x$. This effect is much more pronounced at LHC.

Based on the calculation of modified fragmentation functions, we test the consequences of the two discussed scenarios and compute the nuclear modification factor $R_{AA}$. This experimental observable can be estimated by folding fragmentation functions with the differential cross section $\frac{{\rm d}\sigma}{{\rm
d}q_\perp^2}$ for the production of the fast parton. For our
comparison, we use Pythia 6.4 \cite{Sjostrand:2006za} for a LO
calculation of this cross section in $pp$-collisions at $\sqrt{s}=200$
GeV. We can fit the Pythia output with the form
\begin{equation}
\frac{{\rm d}\sigma}{{\rm d}q_\perp^2}=A\left(1
+\frac{q_\perp^2}{1 \mbox{ GeV}^2}\right)^B \left(1-\frac{2q_\perp}{\sqrt{s}}\right)^C.
\end{equation}
For quark production at RHIC, the parameters have the following values:
\begin{equation}
A \simeq 468.706 \frac{\rm mb}{\rm GeV^{2}}, \qquad B \simeq -3.04, \qquad
C \simeq  9.69.
\end{equation}
For LHC, the cross section for gluon jet production becomes much larger than for quark jets for typical values of $p_\perp$ and
consequently our approximation of a gluonic evolution is
realistic. In the parametrization for $\frac{{\rm d}\sigma}{{\rm d}q_\perp^2}$, 
we change the parameter values for LHC (with $\sqrt{s}=5.5$ TeV) as follows:
\begin{equation}
A \simeq  524395 \frac{\rm mb}{\rm GeV^{2}}, \qquad B \simeq  -3.27,\qquad
C \simeq 10.75.
\end{equation}

\begin{figure}[!t]
\centering \includegraphics[width=0.8\textwidth]{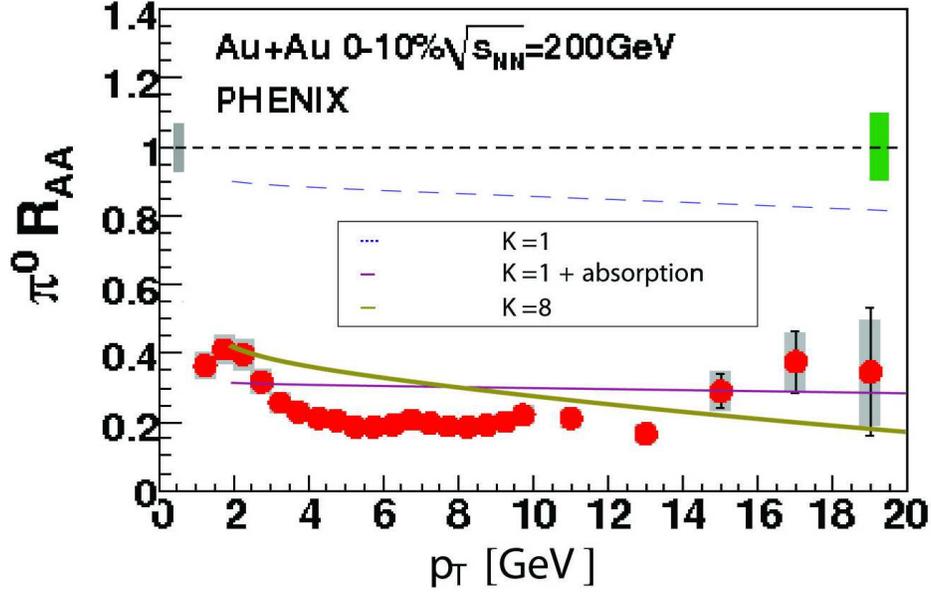}
\caption{\label{RAA-rhic} Two scenarios for the nuclear modification factor $R_{AA}$ of pions at RHIC as a function of their transverse momentum $p_\perp$. The full drawn thin curve corresponds to $K=1$ in \eqnref{Sf} while the full drawn thick curve is calculated for a scenario with $K=8$. The dashed   curve gives $R_{AA}$ without absorption. For comparison, experimental data from PHENIX \cite{Adare:2008qa} are also shown.
}
\end{figure}

Neglecting initial state effects,
the nuclear modification factor can be written as a ratio of the
medium fragmentation function folded with the parton cross section to the vacuum fragmentation function folded with the same cross section.
\begin{equation}\label{RAA_est}
R_{AA}(p_\perp)\simeq \frac{\int \mbox{d}z\,\mbox{d}q_\perp^2
\frac{{\rm d}\sigma}{{\rm d}q_\perp^2} D^{m}(z,Q^2) \delta(z
q_\perp-p_\perp)}{\int \mbox{d}z\,\mbox{d}q_\perp^2 \frac{{\rm
d}\sigma}{{\rm d}q_\perp^2} D^{v}(z,Q^2) \delta(z q_\perp-p_\perp)}
\end{equation}
The results for $R_{AA}$ in our RHIC scenario for $K=8$ (full drawn thick curve) and $K=1$ ( 
dashed curve) are shown in \figref{RAA-rhic}. In this plot, it is clearly
visible that the natural choice $K=1$ cannot account for RHIC data for
$R_{AA}$ (dashed curve), i.e. in scenario 1 parton energy loss alone cannot explain the data.
But an artificial tuning of $K$ -- and consequently of $\hat
q$ -- to $K= 8$  matches the observed nuclear
modification factor. 
 
From our perspective, scenario 1 can compete with scenario 2,
when an additional mechanism for suppression, namely  the
absorption of resonances in the resonance gas is included. This additional 
physics is quite natural, since the lifetime of the plasma and the time for the shower
evolution are comparable in magnitude and smaller than the maximum path length allowed by the size of the 
hot medium. Therefore when the plasma fades out, confinement is no longer an issue and 
the created resonance gas  will allow prehadrons as interaction partners 
with hadronic cross sections. In the next section, we will discuss in detail this
scenario of resonance absorption.

\begin{figure}[!t]
\centering
\includegraphics[width=0.8\textwidth]{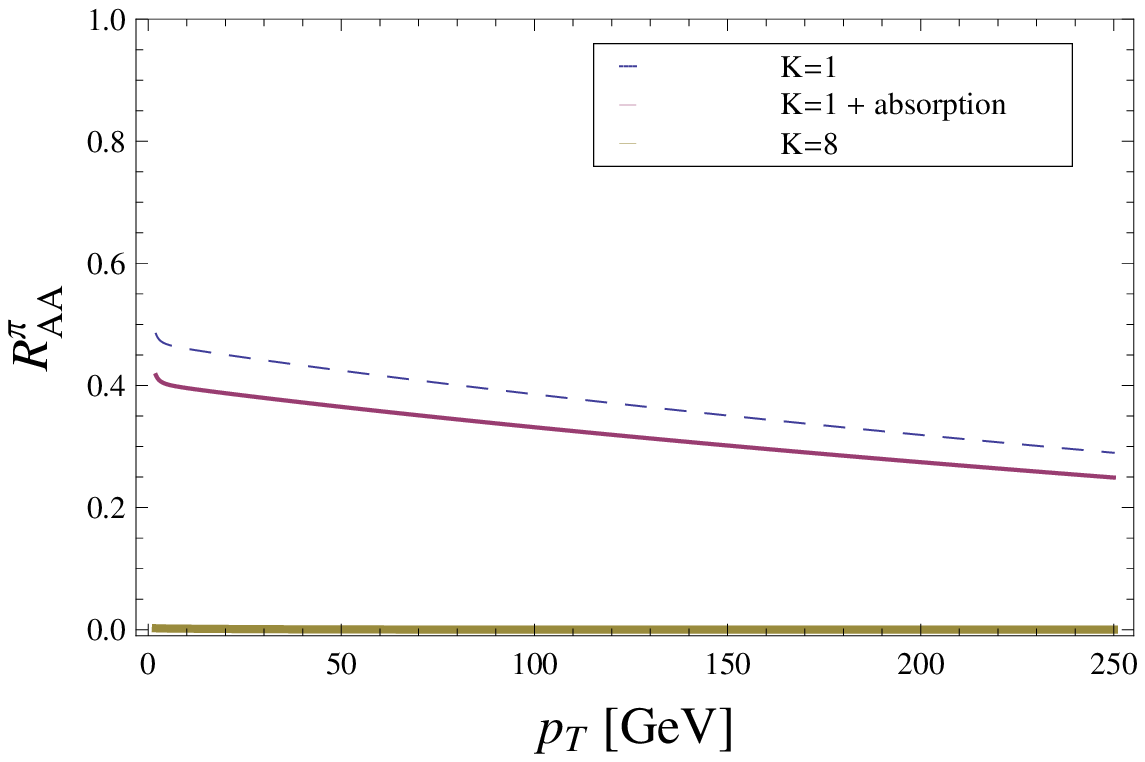}
\caption{\label{RAA-lhc} Two scenarios for the nuclear modification factor $R_{AA}$ of pions at LHC as a function of their transverse momentum $p_\perp$. The  dashed curve corresponds to $K=1$ in \eqnref{Sf} while the full drawn thin curve is calculated for a scenario with $K=8$. The dashed  curve gives $R_{AA}$ without absorption.
}
\end{figure}

For LHC, we also use \eqnref{RAA_est} to compute the nuclear
modification factor for pions.  The result is shown in
\figref{RAA-lhc} for $K=1$ and $K=8$. The larger temperature at LHC
leads to an increase of the suppression relative to RHIC. For the case
of $K=8$, the computed $R_{AA}$ is $10^{-3}-10^{-5}$ which represents a drastic medium effect. Observing such a large effect at LHC would give strong support for a picture in which the suppression is caused by parton energy loss with
a large transport parameter $\hat q$.

The other scenario with $K=1$, however, would tell a different story: As discussed in the next section, we
expect the absorption of resonances to be much less important at LHC
than compared to RHIC. Consequently, the picture of partonic energy
loss becomes more realistic and the calculation with $K=1$ becomes
closer to reality. In this way, the nuclear modification factor could
increase from RHIC to LHC (i.e. less suppression) although the energy
loss increases.


\section{Effect of Prehadron absorption in Hot Resonance Matter}

Various time and length scales are involved
in the evolution from the quark-gluon plasma to the hadronic
phase. Here, we collect numerical values for these time scales 
to argue that absorption plays an important role at RHIC, but is much less relevant at LHC.

We consider the following time scales:
\begin{itemize}
\item Equilibration time $\tau_0$ of the plasma, which is estimated to
be of the order of $0.5$ fm at RHIC but much smaller ($<0.2$ fm) at
LHC \cite{Arsene:2004fa}.
\item Life time of the plasma which we estimate from longitudinal Bjorken expansion
\begin{equation}\label{tauc}
\tau_c=\tau_0 \left(\frac{T_0}{T_c}\right)^3
\end{equation}
For an initial temperature $T_0=0.3$ GeV at RHIC and $T_0=0.5$ GeV at LHC,
we find $\tau_c=3.3$ fm at RHIC and $\tau_c=6.1$ fm at LHC. 
\item  Average time for evolution of the parton from the hard scale $Q_{\rm max}$ to $Q_0$ (where prehadrons can be formed):
\begin{equation}\label{tauevo}
\tau_{\rm evo}=\frac{E}{Q_0^2}-\frac{E}{Q_{\rm max}^2} 
\end{equation}
With $E=Q_{\rm max}=20$ GeV for RHIC and $E=Q_{\rm max}=100$ GeV for LHC (and
$Q_0=\sqrt{2}$ GeV), we find $\tau_{\rm evo}=2$ fm for RHIC and
$\tau_{\rm evo}=10$ fm for LHC. Note that these time estimate apply to hadrons with average $x$-fraction.
\item Time in the hot zone $\tau=R/c$:  For central Au+Au collisions and
central Pb+Pb collisions, the size of the plasma  is almost identical when estimated in terms of the size of the nuclei involved: $R_{\rm Au}=6.9$ fm and $R_{\rm Pb}= 7.1$ fm.
\end{itemize}
The hierarchy between time scales is different at RHIC and LHC. 
For RHIC, we have the following situation: After the evolution of the shower
to the virtuality $Q_0$,  the plasma has
become a resonance gas and preconfined states are formed which can
be absorbed in the resonance gas. For an estimate of this effect, we use an exponential form for the nuclear attenuation, averaged over production
points ${\bf x}_0$ and path-length (determined by its angle $\phi_l$ of inclination) in a
2-dimensional pancake-like system,
\begin{equation}\label{rabs}
r_{\rm abs}=\frac{1}{2\pi}\frac{1}{\pi R^2}\int \mbox{d}^2 {\bf x}_0\,
\mbox{d}\phi_l\frac12 \left(e^{-n_{res} \sigma_{\rm res}
l_1(R,x_0,\phi_l)}+e^{-n_{\rm res} \sigma_{\rm res}
l_2(R,x_0,\phi_l)}\right).
\end{equation}
In this equation, $l_1$ and $l_2$ are the paths of the two partons which
are flying in opposite directions in a medium of size $R_{\rm Au}$ or $R_{\rm Pb}$
respectively. The two respective path lengths are determined by $|{\bf x}_0\pm (l_i + \tau_c){\bf e}_l|=R$.  For the
absorption cross section, we use estimates about the resonance gas. These excited hadrons have to deconfine and consequently, their size has to be larger than in vacuum. 
Therefore, the hadronic cross section is estimated to be larger than the $\pi\pi$
cross section (and similar to the $\pi N$ cross section) because of the increased size of the resonances.
\begin{equation}\label{sigmares}
\sigma_{\rm res}\simeq 30 \mbox{ mb}
\end{equation}

The above cross section concerns hot matter. In cold matter, the medium consists of unexcited nucleons. Therefore, the formation of the prehadron plays a more important role in the determination of the prehadron-nucleon cross section. Indeed, a cross section reduced by a factor of 2/3 compared to its vacuum value fits the suppression rates of hadrons in deep-inelastic scattering on nuclei \cite{Accardi:2005jd}.

For the density of resonances, we take $n_{\rm res} \approx T_c^3$ \cite{Gerber:1988tt}.
The duration of this resonance phase is considerably longer than the plasma phase,
since the entropy density has to be dumped into a large volume. 

With the estimate given in \eqnref{sigmares}, one finds values of $r_{\rm abs}\simeq 0.35$. The final result for the nuclear modification factor, $R_{AA}^{\rm tot}=R_{AA}\, r_{\rm abs}$, is shown in \figref{RAA-rhic} (full drawn thin curve).
This scenario of combined suppression mechanisms can principally account for the large suppression observed in the
nuclear modification factor $R_{AA}$ at RHIC.

The situation at LHC is, however, qualitatively different. For a 100 GeV jet, the development of the parton shower takes about 10 fm and we have $\tau_c< \tau_{\rm evo}$. Consequently, the parton shower will typically not be surrounded by a deconfined medium already at virtuality $Q_1>Q_0$. Since $\tau_c \simeq R$, we can estimate an average value for $Q_1$ by demanding
\begin{equation}
 R=\frac{E}{Q_1^2}-\frac{E}{Q_{\rm max}^2} .
\end{equation}
The finite size of the medium can be included by defining two cut-off
scales: At $Q_1>Q_0$ scatterings stop and at $Q_0$ perturbative splittings
stop. For $Q_0<Q<Q_1$, we use vacuum evolution equations.

A finite medium
length reduces the suppression factor slightly. For a medium of length $R=7$ fm, the nuclear modification factor roughly increases from 0.4 to 0.6 roughly in the scenario with $K=1$. For the second scenario with $K=8$ (large $\hat q$), we find $R_{AA}\simeq 10^{-2}-10^{-3}$ instead of $10^{-3}-10^{-5}$.

After the evolution to $Q_0$, we may also have resonance absorption at LHC. But since $\tau_{\rm evo}\simeq$10 fm is not much smaller than 2$R$, absorption cannot play a major role. If we let $\tau_{c}$ take the role of $\tau_{\rm evo}$ in \eqnref{rabs}, we find a value of $r_{\rm abs}\simeq0.86$ from this equation.

The final result for scenario 1 ($K=1$ including absorption) for LHC is shown in \figref{RAA-lhc} (full drawn thin curve). Clearly, absorption is not very important at LHC.


\section{Modified Parton Distribution in a Jet at small
  $x$}\label{sec:gaussmed}

In this section, we study the quark/gluon composition of the jet for RHIC/LHC. At small $x$, the presence of the dense and hot
medium modifies the peak position of the Hump-Backed Plateau in the $\ln(1/x)$-distribution. As in the previous section, we consider a quark evolution equation in a scenario for RHIC and a gluonic evolution equation for LHC. Here, we use the same scattering
term as in the previous section.
For the evolution at small $x$, a key ingredient is soft gluon
coherence. Without soft gluon coherence, the parton distribution would
diverge at small $x$.
In order to suppress gluon radiation at small $x$,  we solve the
evolution equation with the scale $z^2 Q^2$ in the parton distribution
\cite{Ellis}
\begin{equation}
 \frac{ \partial D(x,Q^2)}{\partial\ln Q^2} =\frac{\alpha_s(Q^2)}{2
\pi} \int_{x}^1 \frac{\mbox{d}z}{z} P(z)
D\left(\frac{x}{z},z^2Q^2\right)+S(x,Q^2).
\end{equation}
For small $x$, we can reduce the splitting function to $\tilde P(z)= 2 C_A/z$ for gluons (LHC) and $\tilde P(z)= 2 C_F/z$ for quarks (RHIC) since these terms represent the dominant contribution. In the following, we use the splitting function $ P(z)= 2 C_R/z$ to account for both cases. Such an approach is related to the double-logarithmic approximation (DLA) \cite{dokshitzer:1991} of parton eolution. More terms are taken into account in the framework of MLLA and NMLLA \cite{PerezRamos:2007cr,Ramos:2006mk}.

The scattering term $S(x,Q^2)$ has the same form as in section 2. Its construction and relation to the jet transport parameter $\hat q$ is discussed in \eqnref{Sterm}.
Since the thermal gluon mass $m_s\propto T$, the scattering term is proportional to $T^2/Q^2$ which makes it higher twist, a feature shared with the formalisms developed in \cite{Wang:2001ifa,Qiu:1990xxa}. Here, we rewrite the scattering term as
\begin{equation}\label{S_der}
S(x,Q^2)= \epsilon  \alpha_s^2(Q^2) \frac{T^2}{Q^2}
\left(D(x, Q^2)+x\frac {\partial D(x,Q^2)}{\partial x} \right)
\end{equation}
where
\begin{equation} \label{epsilon}
\epsilon     = \left\{
\begin{array}{cc}
5.4\, K& \mbox{quark jet}\\
12.2\, K & \mbox{gluon jet}
\end{array}\right.
\end{equation}
is a dimensionless constant used as an abbreviation for the prefactors. The numerical values are calculated with the help of \eqnref{qhatval}.

The parameter $K$ allows us to study the two scenarios discussed in the previous section: Scenario 1
($K=1$) includes small parton energy loss and resonance
absorption. Scenario 2 ($K=8$) parametrizes a large transport parameter
$\hat q$ tuned up to fit the RHIC data with parton energy
loss alone.  For the calculation of the parton distribution at small
$x$ it is crucial to keep track of the dependence on $\alpha_s$.  
We study the most essential modifications and
solve the evolution equation in Gaussian approximation
\cite{Ellis}. This method reproduces experimental
data quite well. The
dependence of the scattering term on $\alpha_s^2$ 
and the $x$-dependence of the scattering kernel preserves the jet multiplicity,
but  modifies the position of the maximum and the width of the parton distribution
in $\ln(1/x)$. For further analysis,  we define the Mellin transform of $D(x,Q^2)$  by
\begin{equation}
d(J,Q^2)=\int_0^1 \mbox{d} x\, x^{J-1} D(x, Q^2).
\end{equation}
In Mellin space we can simplify the scattering term by partial
integration:
\begin{equation}
\epsilon \alpha_s^2 \frac{T^2}{Q^2} \int_0^1 \mbox{d} x\, x^{J-1}
\left(D(x,Q^2)+x \frac{\partial D}{\partial x}(x,Q^2)\right)=\epsilon
\alpha_s^2 \frac{T^2}{Q^2} \left(d(J, Q^2)- J d(J, Q^2)\right)
\end{equation}
For constant $\alpha_s$, the evolution equation reads after Mellin
transformation
\begin{equation}\label{ev_gam}
 \frac{ \partial d(J,Q^2)}{\partial \ln Q^2}
=\frac{\alpha_s}{2 \pi}
\int_{0}^1
\mbox{d}u P(u)
d\left(J, u^2Q^2\right)-\epsilon \alpha_s^2 \frac{T^2}{Q^2} (J-1) d(J,Q^2).
\end{equation}
We use the following ansatz for the parton distribution in
Mellin space with the anomalous dimension $\gamma$
\begin{equation}\label{d_as}
d(J, Q^2)\propto \left(\frac{Q^2}{Q_0^2}\right)^{\gamma(J,\, \alpha_s)}
\end{equation}
for which we obtain a consistency equation
$\gamma$
\begin{equation}\label{gamma}
\gamma(J, \alpha_s)=\frac{\alpha_s C_R}{\pi}\frac{1}{J-1+2 \gamma(J,
\alpha_s)}- \epsilon \alpha_s^2 \frac{T^2}{Q^2} (J-1)
\end{equation}
At $J=1$, the contribution from the scattering term vanishes. Since the first moment with
$J=1$ of the parton distribution is the jet multiplicity, the multiplicity will
not change. This result is natural for a
$2\to 2$-scattering term.  Modifications coming from
the scattering term are suppressed since it is 
a higher-twist  $\propto T^2/Q^2$ contribution. Nevertheless it will
turn out that the medium  significantly  modifies the parton distribution $D(x,Q^2)$.

From Eq. (\ref{gamma}) we find for the anomalous dimension:
\begin{eqnarray}\label{gamma2}
\gamma(J, \alpha_s)=-(J-1)\left(\frac14 + \frac{\epsilon
\alpha_s^2}{2} \frac{T^2}{Q^2}\right)+\sqrt{(J-1)^2\left(\frac14 -
\frac{\epsilon \alpha_s^{2}}{2} \frac{T^2}{Q^2}\right)^2+
\frac{\alpha_s C_R}{2\pi}}
\end{eqnarray}
The small-$x$ behaviour of the parton distribution is related to its
low Mellin moments, therefore we can expand the anomalous dimension 
around $J=1$. 
\begin{eqnarray}\label{gamma3}
\gamma(J, \alpha_s) \simeq \sqrt{\frac{\alpha_s
C_R}{2\pi}}-(J-1)\left(\frac14 + \frac{\epsilon \alpha_s^2}{2}
\frac{T^2}{Q^2}\right) +\frac12
\sqrt{\frac{2\pi}{\alpha_s C_R}} (J-1)^2\left(\frac14 - \frac{\epsilon
\alpha_s^2}{2} \frac{T^2}{Q^2}\right)^2
\end{eqnarray}
This expansion preserves the
most essential features of the parton distribution in vacuum. Since the
medium does not change the dominant behavior of the splitting term,
it also works in the medium.
For running $\alpha_s$, we use \cite{Ellis}
\begin{equation}\label{noneikonal}
d(J,Q^2)\propto \exp\left(\int_{Q_0^2}^{Q^2} \frac{\mbox{d}Q'^2}{Q'^2}
\gamma(J, \alpha_s(Q'^2))\right)
\end{equation}
For the calculation of the integral, we use the perturbative 1-loop coupling
\begin{equation}
\alpha_s(Q^2)=\frac{1}{b \ln\left(\frac{Q^2}{\Lambda_{\rm QCD}^2}\right)},\qquad b=\frac{11-\frac{2n_f}{3}}{4\pi},\quad \Lambda_{\rm QCD}=250\mbox{ MeV}.
\end{equation}

The number of flavors is set to five since we study jets with energies well below the mass of the top quark.
A calculation with $\alpha_s$ in one loop approximation 
is accurate enough for the vacuum evolution. 
Note that $d(J,Q^2)$ in \eqnref{noneikonal} has a Gaussian dependence on $(J-1)$ because the
Taylor expansion in \eqnref{gamma3} stops at second order:
\begin{equation}\label{dJ}
d(J,Q^2)=C \exp\left(a_0+a_1 (J-1)+ a_2 (J-1)^2\right).
\end{equation}
The normalization of $d(J,Q^2)$ is not predicted by the ansatz
\eqnref{noneikonal}. 
We fix the normalization $C\simeq 0.024$ of the multiplicity
$n(Q^2)= C \exp(a_0)$ with LEP data from e$^+$e$^-$
\cite{e+e-,Schmelling:1994py}. It is possible to use these ``vacuum''
data since the scattering term leaves the multiplicity unchanged.

The coefficients $a_i$ in \eqnref{dJ} result from the integration in
\eqnref{noneikonal} and depend on both virtuality and
temperature. They are given as follows:
\begin{eqnarray}
a_0&=&\displaystyle\frac1b \sqrt{\frac{2 C_R}{\pi\alpha_s(Q^2)}}-\left[Q^2\to Q_0^2\right]\label{a0med}\\
a_1&=&\frac{1}{4b\alpha_s(Q^2)}-\frac{\epsilon}{2b} \frac{T^2}{Q^2} \alpha_s(Q^2) 
\nonumber\\&& - \frac{\epsilon}{2b^2} \frac{T^2}{\Lambda^2}\,\mbox{Ei}\left(-(b\alpha_s(Q^2))^{-1}\right)-\left[Q^2\to Q_0^2\right]
\nonumber\\ \label{a1med}
a_2&=&\frac{1}{24b}\sqrt{\frac{\pi}{2C_R}}\,\alpha_s(Q^2)^{-3/2}+ \frac{1}{2\sqrt{2}b^2}
\frac{T^2}{\Lambda^2} \frac{\pi}{\sqrt{C_R}}  \mbox{erf}\left((b\alpha_s(Q^2))^{-1/2}\right)\epsilon\nonumber \\
&&-\frac{4}{15\,b^3}\sqrt{\frac{2\pi}{C_R}}\frac{T^4 }{Q^4}\,\alpha_s(Q^2)^{1/2}\epsilon^2
+\frac{1}{2b}\sqrt{\frac{\pi}{2C_R}}\frac{T^2}{Q^2} \alpha_s(Q^2)^{1/2}\epsilon\nonumber\\
&&+\sqrt{\frac{\pi}{2C_R}}\frac{T^4}{Q^4}\,\left(\frac{2}{15}\frac{\alpha_s(Q^2)^{3/2}}{b^2}-\frac{1}{10}\frac{\alpha_s(Q^2)^{5/2}}{b}\right)\epsilon^2\nonumber\\
&&-\frac{8}{15\,b^{7/2}}\frac{T^4}{\Lambda^4}\frac{\pi}{\sqrt{C_R}}\mbox{erf}\left(\sqrt{\frac{2}{b\alpha_s(Q^2)}}\right)\epsilon^2 -\left[Q^2\to Q_0^2\right]\label{a2med}
\end{eqnarray}
Here, the notation $-\left[Q^2\to Q_0^2\right]$ means that one has to
subtract the previous terms, replacing $Q^2$ by $Q_0^2$. This
subtraction originates from the lower limit of integration in
\eqnref{noneikonal}. Ei($-z$) is the exponential integral function,
defined by
\begin{equation}
\mbox{Ei}(-z)=-\int_z^\infty \mbox{d}t \frac{\mbox{e}^{-t}}{t}.
\end{equation}
For a better illustration of our results, we also give the lowest-order expansion of the medium modification of the coefficient in $\alpha_s$ in \tabref{tab1}. 

\renewcommand{\arraystretch}{1.5}
\begin{table} 
\begin{center}
\begin{tabular}{|l||c|c|} \hline
 & vacuum & medium\\ \hline\hline
$a_0$  &$\quad\displaystyle\frac1b \sqrt{\frac{2 C_R}{\pi\alpha_s}}\quad$ & 0 \\ \hline
$a_1$ & $\displaystyle\frac{1}{4 b \alpha_s}$ & $-\frac{\epsilon}{2} \frac{T^2}{Q^2} \alpha_s^2$  \\ \hline
$a_2$ &$\displaystyle\sqrt{\frac{\pi}{2 C_R}}\frac{1}{24 b} \alpha_s^{-3/2}$  
	& $\frac{\epsilon}{4}\sqrt{\frac{\pi}{2C_R}}\frac{T^2}{Q^2} \alpha_s^{3/2}$  \\ \hline
\end{tabular}
\end{center}
\caption{\label{tab1} Leading terms of the Gaussian coefficients $a_0$, $a_1$ and $a_2$ of the parton distribution in a jet in Mellin space, i.e. $d(J, Q^2)\simeq\exp(a_0+a_1 (J-1)+ a_2 (J-1)^2))$, which determine multiplicity ($a_0$), peak position ($a_1$) and width ($a_2$) of the $\ln(1/x)$-distribution. 
The first column gives the value in vacuum while the second column shows the 
correction due to the medium with the lowest power in $\alpha_s$. 
We use a shorthand notation here: One should take $\alpha_s\to\alpha_s(Q^2)$ in every 
term and subtract the same term with $Q^2\to Q_0^2$.}
\end{table}

The medium modification is solely contained
in the modification of the coefficients $a_i$.  The coefficients $a_i$
have the same physical meaning as in vacuum: $a_0$ describes the
$Q^2$-behavior of the jet multiplicity $n(Q^2)=C\exp(a_0)$, $a_1$
gives the peak position of the distribution in ln$(1/x)$ (see below)
and $a_2$ is related to the Gaussian width. The first term for each
coefficient represents the vacuum result while the second terms proportional to
$\epsilon T^2/Q^2$  represent the modified evolution in \tabref{tab1}.


In vacuum, the coefficients have inverse powers of
$\alpha_s(Q^2)$. Therefore the vacuum moments are determined by the
upper virtuality $Q^2$ with $\alpha_s(Q^2) <\alpha_s(Q_0^2)$. The
terms from the medium-modified evolution have positive powers of
$\alpha_s$ and change the vacuum results.   The medium
corrections are most important at the infrared scale $Q_0^2$, where
$\alpha_s(Q_0^2)>\alpha_s(Q^2)$. Therefore, in the medium the peak position is shifted to larger values of $\ln(1/x)$ while the width decreases.

The well-known MLLA contribution to the peak
position, 
\begin{equation}
\Delta a_1^{\rm MLLA}=\frac{1}{2b}\left(\frac{11N_c}{3}+\frac{2n_f}{3N_c^2}\right)\,\frac{1}{\sqrt{32N_c\pi}}\,\alpha_s(Q^2)^{-1/2},
\end{equation}
cannot be
deduced from   the Gaussian approximation and consequently is added by hand. This contribution is subleading but highly relevant for comparison to experimental data. 

Finally, the parton distribution can be calculated by an
inverse Mellin transformation.
\begin{eqnarray}
D(x,Q^2)&=&\frac{C}{2 \pi i x}\int_{1-i \infty}^{1+i \infty} dJ \frac{1}{x^{J-1}}
\exp( a_0  -a_1 (J-1)+ a_2 (J-1)^2 ) \nonumber\\
        &=& \frac{C}{2 \pi x}\int_{-\infty}^{+\infty} d\tilde J
\exp\left[a_0 +i \tilde J \left( -a_1+\ln\left[\frac1x\right]\right)- a_2 \tilde J^2  \right]
\end{eqnarray}
Performing the Gaussian integral yields
\begin{equation}
x\,D(x, Q^2)= \frac{n(Q^2)}{2\sqrt{\pi a_2}} \exp\left(-\frac{\left(\ln(\frac1x)-a_1\right)^2}{4 a_2}\right).
\end{equation}

\begin{figure}
\begin{center}
\epsfig{file=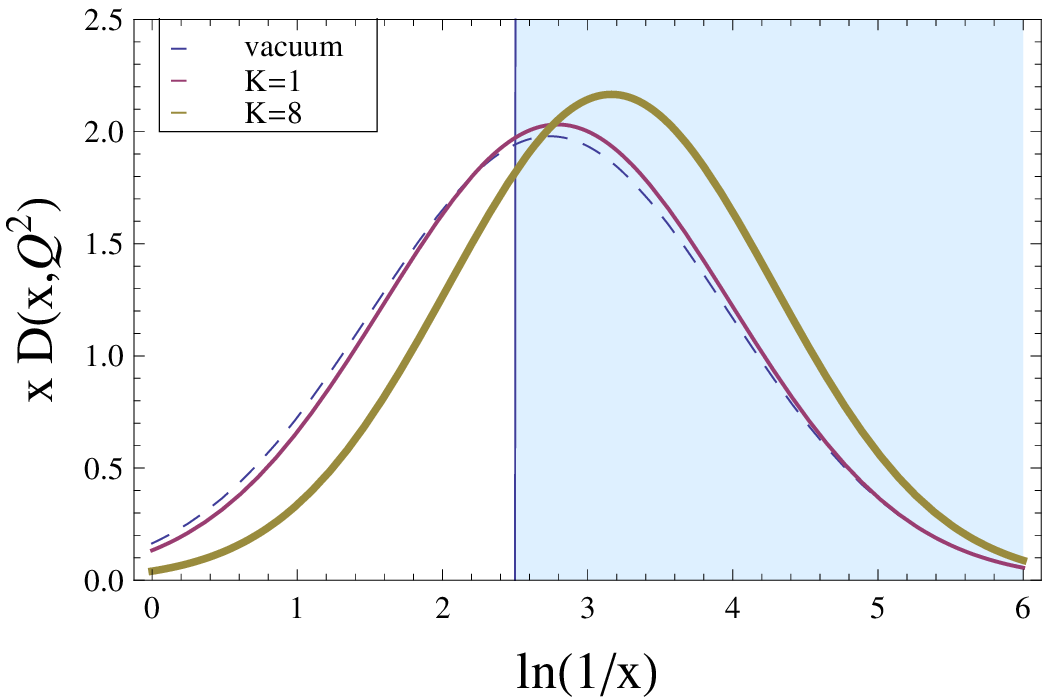,width=0.8\textwidth} \epsfig{file=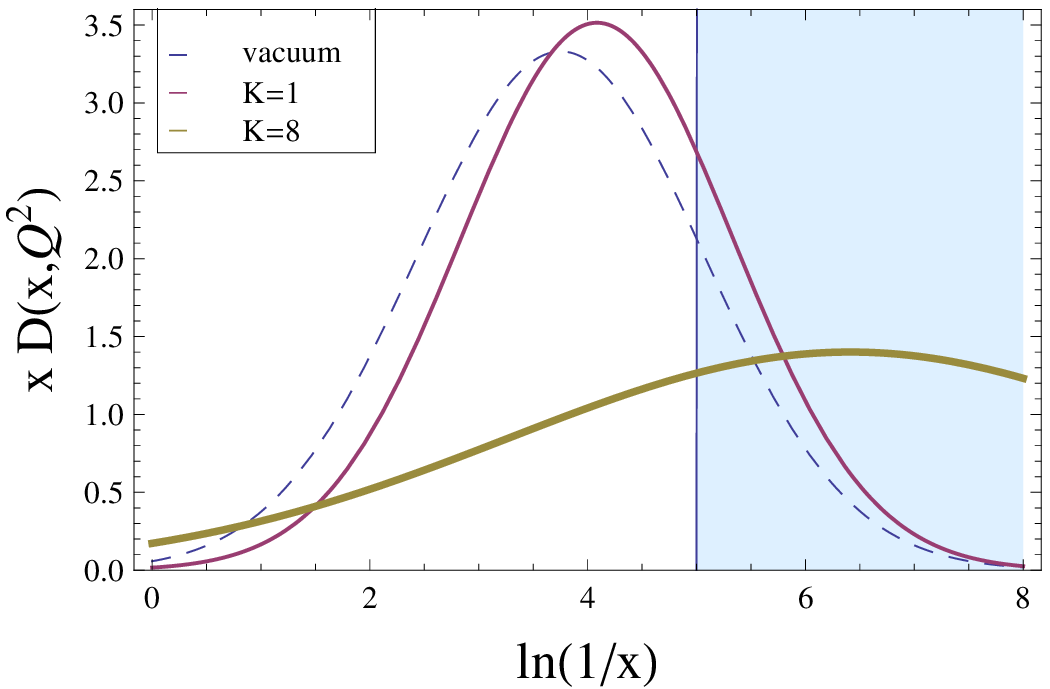,width=0.8\textwidth}
\caption{\label{xD} Top: Quark distribution $xD(x, Q^2)$ in a jet in
  Gaussian approximation shown for $Q^2=(20\mbox{ GeV})^2$ for
  evolution in vacuum (  dashed curve) and medium-modified evolution
  with $K=1$ (full drawn thin curve) and $K=8$ (full drawn thick curve)
  in a plasma with temperature $T=0.3\mbox{ GeV}$ as a function of
  $\ln(1/x)$. Bottom: Same for gluons with $Q=100$ GeV and $T=0.5$ GeV.  }
  \end{center}
\end{figure}

In Fig.~\ref{xD} we show a comparison of the logarithmic parton
distributions $xD(x,Q^2)$ in a jet in vacuum and medium. As discussed after \eqnref{epsilon}, we use the values $K=1$ and $K=8$ and compare to the vacuum results. Two different kinematical cases are studied: $Q_{\rm
  max}=20$ GeV and $T=0.3$ GeV, suitable for RHIC (top), and $Q_{\rm
  max}=100$ GeV and $T=0.5$ GeV, relevant for LHC, in the bottom. 
The shaded area in both figures represents estimates for the region where particles cannot be distinguished from the plasma anymore
\cite{Bruna:2009kw,Estienne:2009si}.

At RHIC, the scattering term basically shifts the peak
position of the quark distribution to larger $\ln(1/x)$ (i.e. smaller $x$) due to the energy loss 
from  gluon scattering with plasma gluons. The modification of the
parton distribution in scenario 1 with $K=1$ is rather small. In
contrast, the parton distribution in scenario 2 with $K=8$ differs
enormously from the vacuum one. Such a strong modification has not been observed
in (preliminary) experimental analyses \cite{Bruna:2009kw}.
Note that for jets the existence of the resonance matter is unimportant.

For LHC, particle identification above the soft background from the
plasma is possible over a much wider range. Scenario 2  with $K=8$ leads to a strong modification of the gluon
distribution: The peak position is shifted to larger values of
$\ln(1/x)$ and the distribution becomes broader. In contrast, scenario 1 with
$K=1$ again leads only to small changes compared to vacuum.

\section{Summary and discussion}
We have presented two scenarios for jet evolution in nucleus-nucleus collisions. Scenario~1 considers weak parton energy loss caused by scattering in the plasma. After the evolution to a hadronic scale $Q_0$, preconfinement causes hadronic absorption in the resonance gas which is left directly after the plasma decay. In contrast, scenario 2 is solely based on strong parton energy loss. In both scenarios, the partonic evolution is computed in the framework of a modified DGLAP equation which combines radiation and scattering of the fast parton in the medium. The corresponding scattering term is of higher twist, $S(x,Q^2) \propto T^2/Q^2$.

In scenario 1, we estimate jet transport parameters of $\hat q\simeq 0.5$ GeV$^2$/fm at RHIC and $\hat q\simeq 5.2$ GeV$^2$/fm at LHC. Both parameters have to be understood as average values since we do not consider the change of $\hat q$ with decreasing temperature and density of the plasma. Enhancement of the QCD estimate $\hat q\simeq 0.5$ GeV$^2$/fm by a $K$-factor $K=8$ in scenario 2 can also accomodate our model to the RHIC data for $R_{AA}$. Such an enhanced value may correspond to the hypothesis of a strongly interacting quark-gluon plasma (sQGP).

We do not favour scenario 2, since there are strong evidences for the importance of absorption: (i)
Scenario 2 fails to describe production of leading hadrons in
deep-inelastic scattering \cite{wang2},
while models including absorption do well \cite{knp1,knp2,Accardi:2005jd}; (ii)
 $J/\Psi$ suppression in the hot medium can be described with a rather small value of the transport coefficient
\cite{Kopeliovich:2010vk}, which agrees well with the pQCD expectations; (iii) Even with
the enhanced transport coefficient, scenario 2 does not explain the
observed strong suppression of open beauty \cite{dgw},
while the effects of absorption make suppression of light and heavy quarks
similar \cite{paradigms}.

In contrast, we argue that due to the shorter life times of the shower and the plasma at RHIC (compared to the extension of the nuclei), the parton can become preconfined during its way through the medium and interact as a prehadron with the decaying resonance matter. For an estimate of this effect, we use a standard initial density $n_{\rm res}\approx T_c^3$ of resonances and an interaction cross section of $\sigma_{\rm res}=30$ mb. With these values, we obtain an acceptable fit
 to the RHIC data with $K=1$.

The new energy domain of LHC allows to test these two alternatives in two different ways. First, we think that due to the similarity of $\alpha_s(\pi T)$ and $\alpha_s(Q_0)$ for $T=500 $ MeV, $K=1$ is also reasonable for LHC. Then, we predict in scenario~1 a slightly larger $R_{AA}$ due to a stronger partonic energy loss while hadronic absorption plays a minor role. At LHC, the length of the perturbative shower is much longer, leaving almost no time for resonance absorption in the medium after preconfinement. Consequently, prehadron absorption does not increase the suppression factor significantly at LHC. In contrast to scenario~1, scenario~2 would lead to a very large suppression, i.e. $R_{AA}<10^{-2}$ at LHC. Such a strong suppression may be detected within the first few months of the heavy-ion program at LHC.

Secondly, one can study the distribution of low-$x$ partons in the jet, since at LHC particles up to $\ln(1/x)\simeq 5$ are measurable above the plasma background. The shape of the hump-backed plateau of the medium-modified jet will clearly show whether any enhancement of $\hat q$ in our calculation is necessary.



For our considerations, we have not made use of further experimental data \cite{Aggarwal:2007gw,:2008cx,Adler:2006bw} for several reasons. First, the results from SPS in Ref.~\cite{Aggarwal:2007gw}  and RHIC in Ref.~\cite{:2008cx} are not fully consistent with each other. The absence of suppression at small energies in \cite{:2008cx}  is unexpected in our absorption scenario. However,
Ref.~\cite{Aggarwal:2007gw} finds a suppression at a similar energy in the most central collisions. Furthermore, we mention as an aside that the suppression is estimated to set in after a propagation length of 2 fm in \cite{Adler:2006bw} which may be related to our estimate for the shower time.


As a final remark, we note that the jet composition in the medium from our calculation can easily be distinguished from approaches with medium-induced gluon radiation \cite{Borghini:2005em}. The enhancement of the splitting functions in these calculations always leads to a much higher peak and consequently to a much larger multiplicity \cite{Ramos:2008qb}.


\vspace{1cm}
\noindent{\large\bf Acknowledgements}\\
The authors are indebted to K.~Reygers for useful discussions on experimental data. 
S.D. has been supported by the Bundesministerium f\"ur Bildung and Forschung (BMBF)
under grant number BMBF 06 HD 196 and 
within the framework of the Excellence Initiative by the
German Research Foundation (DFG) through the Heidelberg Graduate School of
Fundamental Physics (grant number GSC 129/1). This work was supported in part by Fondecyt (Chile) grant 1090291, by DFG
(Germany) grant PI182/3-1, and by Conicyt-DFG grant No. 084-2009.


\begin{thebibliography}{100}




\bibitem{Adler:2003qi}
  S.~S.~Adler {\it et al.}  [PHENIX Collaboration],
  Phys.\ Rev.\ Lett.\  {\bf 91} (2003) 072301
  [arXiv:nucl-ex/0304022].\\
  J.~Adams {\it et al.}  [STAR Collaboration],
  Phys.\ Rev.\ Lett.\  {\bf 91}, 172302 (2003)
  [arXiv:nucl-ex/0305015].


\bibitem{Arsene:2004fa}
  I.~Arsene {\it et al.}  [BRAHMS Collaboration],
  Nucl.\ Phys.\ A {\bf 757} (2005) 1
  [arXiv:nucl-ex/0410020].\\
  B.~B.~Back {\it et al.},
  Nucl.\ Phys.\ A {\bf 757} (2005) 28
  [arXiv:nucl-ex/0410022].\\
  J.~Adams {\it et al.}  [STAR Collaboration],
  Nucl.\ Phys.\ A {\bf 757} (2005) 102
  [arXiv:nucl-ex/0501009].\\
  K.~Adcox {\it et al.}  [PHENIX Collaboration],
  Nucl.\ Phys.\ A {\bf 757} (2005) 184
  [arXiv:nucl-ex/0410003].


\bibitem{Majumder:2010qh}
  A.~Majumder and M.~Van Leeuwen,
  arXiv:1002.2206 [hep-ph].\\
  U.~A.~Wiedemann,
  arXiv:0908.2306 [hep-ph].\\
  C.~A.~Salgado,
  arXiv:0907.1219 [hep-ph].\\
  D.~d'Enterria,
  arXiv:0902.2011 [nucl-ex].

\bibitem{Adare:2006nq}
  A.~Adare {\it et al.}  [PHENIX Collaboration],
  Phys.\ Rev.\ Lett.\  {\bf 98} (2007) 172301
  [arXiv:nucl-ex/0611018].\\
  B.~I.~Abelev {\it et al.}  [STAR Collaboration],
  Phys.\ Rev.\ Lett.\  {\bf 98} (2007) 192301
  [arXiv:nucl-ex/0607012].




\bibitem{Baier:1996kr}
  R.~Baier, Y.~L.~Dokshitzer, A.~H.~Mueller, S.~Peigne and D.~Schiff,
  Nucl.\ Phys.\ B {\bf 483} (1997) 291
  [arXiv:hep-ph/9607355]; 

\bibitem{wang}
X.N. Wang, Phys. Lett. B {\bf 595} (2004), p. 165\\
Q. Wang and X.N. Wang, Phys. Rev. C {\bf 71} (2005), p. 014903

\bibitem{gyulassy}
M. Gyulassy, P. Levai and I. Vitev, Nucl. Phys. B {\bf 594} (2001), p. 371\\
M. Gyulassy, P. Levai and I. Vitev, Phys. Rev. Lett. {\bf 85} (2000), p. 5535\\
M. Gyulassy, P. Levai and I. Vitev, Nucl. Phys. B {\bf 571} (2000), p. 197

\bibitem{wiedemann}
U.A. Wiedemann, Nucl. Phys. B {\bf 588} (2000), p. 303\\
C.A. Salgado and U.A. Wiedemann, Phys. Rev. D {\bf 68} (2003), p. 014008

\bibitem{Zakharov:1998sv}
  B.~G.~Zakharov,
  Phys.\ Atom.\ Nucl.\  {\bf 61} (1998) 838
  [Yad.\ Fiz.\  {\bf 61} (1998) 924]
  [arXiv:hep-ph/9807540].

\bibitem{Shuryak:2004cy}
  E.~V.~Shuryak,
  Nucl.\ Phys.\  A {\bf 750} (2005) 64
  [arXiv:hep-ph/0405066].





\bibitem{Petreczky:2007bn}
  P.~Petreczky,
  Eur.\ Phys.\ J.\ ST {\bf 155} (2008) 123
  [arXiv:0711.2280 [hep-lat]].





\bibitem{Gerber:1988tt}
  P.~Gerber and H.~Leutwyler,
  Nucl.\ Phys.\  B {\bf 321} (1989) 387.

\bibitem{DGLAP}
  L.~N.~Lipatov,
  Sov.\ J.\ Nucl.\ Phys.\  {\bf 20} (1975) 94
  [Yad.\ Fiz.\  {\bf 20} (1974) 181].\\
  V.~N.~Gribov and L.~N.~Lipatov,
  Sov.\ J.\ Nucl.\ Phys.\  {\bf 15} (1972) 675
  [Yad.\ Fiz.\  {\bf 15} (1972) 1218].\\
  G.~Altarelli and G.~Parisi,
  Nucl.\ Phys.\  B {\bf 126}, 298 (1977).\\
  Y.~L.~Dokshitzer,
  Sov.\ Phys.\ JETP {\bf 46}, 641 (1977)
  [Zh.\ Eksp.\ Teor.\ Fiz.\  {\bf 73}, 1216 (1977)].

\bibitem{Albino:2005me}
  S.~Albino, B.~A.~Kniehl and G.~Kramer,
  Nucl.\ Phys.\  B {\bf 725} (2005) 181
  [arXiv:hep-ph/0502188].

\bibitem{Sjostrand:2006za}
  T.~Sjostrand, S.~Mrenna and P.~Z.~Skands,
  JHEP {\bf 0605} (2006) 026
  [arXiv:hep-ph/0603175].

\bibitem{kapusta}
J.~Kapusta, Finite-temperature field theory, Cambridge University Press, 1994



\bibitem{Kopeliovich:2007dt}
  B.~Z.~Kopeliovich, I.~K.~Potashnikova and I.~Schmidt,
 arXiv:0707.4302 [nucl-th].


\bibitem{Bass:2008rv}
  S.~A.~Bass, C.~Gale, A.~Majumder, C.~Nonaka, G.~Y.~Qin, T.~Renk and J.~Ruppert,
  Phys.\ Rev.\  C {\bf 79} (2009) 024901
  [arXiv:0808.0908 [nucl-th]].

\bibitem{CasalderreySolana:2007sw}
  J.~Casalderrey-Solana and X.~N.~Wang,
  Phys.\ Rev.\  C {\bf 77} (2008) 024902
  [arXiv:0705.1352 [hep-ph]].

\bibitem{Adare:2008qa}
  A.~Adare {\it et al.}  [PHENIX Collaboration],
  Phys.\ Rev.\ Lett.\  {\bf 101}, 232301 (2008)
  [arXiv:0801.4020 [nucl-ex]].

\bibitem{wang2} W.~t.~Deng and X.~N.~Wang,
 Phys.\ Rev.\  C {\bf 81}, 024902 (2010)
 [arXiv:0910.3403 [hep-ph]].

\bibitem{Qiu:1990xxa}
  J.~w.~Qiu and G.~F.~Sterman,
  Nucl.\ Phys.\  B {\bf 353} (1991) 105.\\
  J.~w.~Qiu and G.~F.~Sterman,
  Nucl.\ Phys.\  B {\bf 353} (1991) 137.

\bibitem{Bruna:2009kw}
  E.~Bruna  [STAR Collaboration],
  arXiv:0905.4763 [nucl-ex].

\bibitem{Estienne:2009si}
  M.~Estienne  [ALICE Collaboration],
  arXiv:0910.2482 [nucl-ex].


\bibitem{Ellis} 
R.K.~Ellis, W.J.~Stirling and
B.R.~Webber, QCD and Collider Physics, Cambridge University Press 1996.

\bibitem{Karsch:2003vd}
  F.~Karsch, K.~Redlich and A.~Tawfik,
  Eur.\ Phys.\ J.\  C {\bf 29}, 549 (2003)
  [arXiv:hep-ph/0303108].

\bibitem{e+e-}
  W.~Braunschweig {\it et al.}  [TASSO Collaboration],
  Z.\ Phys.\  C {\bf 47} (1990) 187.\\
 W.~Braunschweig {\it et al.}  [TASSO Collaboration],
  Z.\ Phys.\  C {\bf 45} (1989) 193.\\
  D.~Decamp {\it et al.}  [ALEPH Collaboration],
  Phys.\ Lett.\  B {\bf 273} (1991) 181.\\
  P.~Abreu {\it et al.}  [DELPHI Collaboration],
  Z.\ Phys.\  C {\bf 50} (1991) 185.\\
  M.~Z.~Akrawy {\it et al.}  [OPAL Collaboration],
  Phys.\ Lett.\  B {\bf 247} (1990) 617.\\
  B.~Adeva {\it et al.}  [L3 Collaboration],
  Phys.\ Lett.\  B {\bf 259} (1991) 199.

\bibitem{Schmelling:1994py}
  M.~Schmelling,
  Phys.\ Scripta {\bf 51} (1995) 683.


\bibitem{dokshitzer:1991}
Yu.~Dokshitzer, V.A.~Khoze, Basics of perturbative QCD, Edition Frontieres 1991



\bibitem{PerezRamos:2007cr}
  R.~Perez-Ramos, F.~Arleo and B.~Machet,
  Phys.\ Rev.\  D {\bf 78} (2008) 014019
  [arXiv:0712.2212 [hep-ph]].

\bibitem{Ramos:2006mk}
  R.~P.~Ramos,
  JHEP {\bf 0609} (2006) 014
  [arXiv:hep-ph/0607223].\\
	R.~P.~Ramos,
  JHEP {\bf 0606} (2006) 019
  [arXiv:hep-ph/0605083].





\bibitem{Braun:2006vd}
  J.~Braun and H.~J.~Pirner,
  Phys.\ Rev.\  D {\bf 75}, 054031 (2007)
  [arXiv:hep-ph/0610331].




\bibitem{Wang:2001ifa}
  X.~N.~Wang and X.~f.~Guo,
  Nucl.\ Phys.\  A {\bf 696} (2001) 788
  [arXiv:hep-ph/0102230].



\bibitem{Maezawa:2009nd}
  Y.~Maezawa {\it et al.}  [WHOT-QCD Collaboration],
  Nucl.\ Phys.\  A {\bf 830} (2009) 247C
  [arXiv:0907.4203 [hep-lat]].





\bibitem{Domdey:2008gp}
  S.~Domdey, G.~Ingelman, H.~J.~Pirner, J.~Rathsman, J.~Stachel and K.~Zapp,
  Nucl.\ Phys.\  A {\bf 808} (2008) 178
  [arXiv:0802.3282 [hep-ph]].\\
  S.~Domdey, PhD thesis



\bibitem{Arleo:2008dn}
  F.~Arleo,
  Eur.\ Phys.\ J.\  C {\bf 61} (2009) 603
  [arXiv:0810.1193 [hep-ph]].


\bibitem{knp1} B.Z. Kopeliovich, J. Nemchik and E. Predazzi,
Proceedings of the workshop on Future Physics at HERA, ed. by G. Ingelman,
A. De Roeck and R. Klanner, DESY 1995/1996, v. 2, 1038; arXiv:
nucl-th/9607036.

\bibitem{knp2}  B.~Z.~Kopeliovich, J.~Nemchik, E.~Predazzi and
A.~Hayashigaki,
 Nucl.\ Phys.\  A {\bf 740}, 211 (2004).


\bibitem{dgw}  M.~Djordjevic, M.~Gyulassy and S.~Wicks,
 Phys.\ Rev.\ Lett.\  {\bf 94}, 112301 (2005)
 [arXiv:hep-ph/0410372].



\bibitem{Accardi:2005jd}
  A.~Accardi, D.~Grunewald, V.~Muccifora and H.~J.~Pirner,
  Nucl.\ Phys.\  A {\bf 761} (2005) 67
  [arXiv:hep-ph/0502072].\\
  A.~Accardi, V.~Muccifora and H.~J.~Pirner,
  Nucl.\ Phys.\  A {\bf 720} (2003) 131
  [arXiv:nucl-th/0211011].

\bibitem{paradigms}
  B.~Z.~Kopeliovich and J.~Nemchik,
  arXiv:1009.1162 [hep-ph].

\bibitem{Kopeliovich:2010vk}
  B.~Z.~Kopeliovich, I.~K.~Potashnikova and I.~Schmidt,
  Phys.\ Rev.\  C {\bf 82} (2010) 024901
  [arXiv:1006.3042 [nucl-th]].



\bibitem{Aggarwal:2007gw}
  M.~M.~Aggarwal {\it et al.}  [WA98 Collaboration],
  Phys.\ Rev.\ Lett.\  {\bf 100}, 242301 (2008)
  [arXiv:0708.2630 [nucl-ex]].

\bibitem{:2008cx}
  A.~Adare {\it et al.}  [PHENIX Collaboration],
  Phys.\ Rev.\ Lett.\  {\bf 101}, 162301 (2008)
  [arXiv:0801.4555 [nucl-ex]].

\bibitem{Adler:2006bw}
  S.~S.~Adler {\it et al.}  [PHENIX Collaboration],
  Phys.\ Rev.\  C {\bf 76}, 034904 (2007)
  [arXiv:nucl-ex/0611007].


\bibitem{Borghini:2005em}
  N.~Borghini and U.~A.~Wiedemann,
  arXiv:hep-ph/0506218.

\bibitem{Ramos:2008qb}
  R.~P.~Ramos,
  Eur.\ Phys.\ J.\  C {\bf 62} (2009) 541
  [arXiv:0811.2418 [hep-ph]].





\end{thebibliography}
\end{document}